\documentclass[aps,prb,preprint,graphicx,amsbsy,amssymb]{revtex4}

\usepackage{graphicx}
\usepackage{graphics}
\usepackage{epsfig}

\def\be{\begin{equation}}
\def\ee{\end{equation}}
\def\bea{\begin{eqnarray}}
\def\eea{\end{eqnarray}}

\begin{document}
\title{Broken time-reversal symmetry in Josephson junction involving two-band superconductors}
\author{T.K. Ng$^1$, and N. Nagaosa$^{2,3}$}
\affiliation{ $^1$Department of Physics, Hong-Kong University of
Science and Technology, Kawloon, Hong Kong Peoples R China \\
$^2$ Department of Applied Physics, The University of Tokyo, 7-3-1 Hongo, Bunkyo-ku, Tokyo 113-8656, Japan \\
$^3$Cross Correlated Materials Research Group (CMRG), ASI, RIKEN,
WAKO 351-0198, Japan}

\begin{abstract}
 A novel time-reversal symmetry breaking state is found theoretically in the
Josephson junction between the two-gap superconductor and the
conventional s-wave superconductor. This occurs due to the
frustration between the three order parameters analogous to the
two antiferromagnetically coupled XY-spins put under a magnetic
field. This leads to the interface states with the energies inside
the superconducting gap. Possible experimental observations of
this state with broken time-reversal symmetry are discussed.
\end{abstract}


 \maketitle The complex behaviors of the superconductors beyond the BCS theory are now a focus of condensed matter
 physics. Among them, the superconductivity characterized by more than one order parameters, i.e., the multi-gap
 superconductors, is an intriguing and hot topic. Historically, the inter-band mechanism of the pairing for such a
 case has been proposed long ago \cite{Suhl}. A representative example of the multi-gap superconductor is
 MgB$_2$\cite{MgB2} where the specific heat\cite{MgB2heat}, the tunneling\cite{MgB2tunnel}, and angle-resolved
 photoemission spectroscopy (ARPES)
\cite{MgB2ARPES} have revealed the different gap energies for $\sigma$- and $\pi$-bands. The magnitudes of
 the gaps were analyzed by the first-principles band structure calculation \cite{MgB2band} and are found to be
 strongly momentum- and band- dependent. This two-gap behavior is attributed to the strong coupling of the
 $\sigma$-band to the bond stretching phonon mode \cite{ MgB2band}.

  The multi-gap structure should be common in the superconductors with the orbital degeneracy and/or the many
 electron/hole pockets, which is the case for the newly found high temperature superconductor iron
 pnictides \cite{Kamihara}. In these compounds, there are two small electron pockets around M-points and two hole
 pockets around $\Gamma$-point \cite{Singh}. There are many proposals for the gap pairing symmetry\cite{DHLee}, and
 one possibility is that a full gap opens for each pocket, which is consistent with the recent ARPES in Ba$_{0.6}$K$_{0.4}$Fe$_2$\cite{Ding}, although ARPES cannot determine the
 relative sign of the order parameters. Therefore, the determination of the relative sign of the order parameters
 on the pockets is now an important issue to fix the microscopic mechanism for the superconductivity. A clue has
 been given by the resonant magnetic scattering \cite{neutron} which is attributed to the triplet
 exciton in the superconducting state. A comparison with the earlier theoretical analysis \cite{Flude} suggests
 that the relative sign of the order parameter is minus in Ba$_{0.6}$K$_{0.4}$Fe$_2$.

 In this paper, we explore theoretically a novel phenomenon in the two-gap superconductors when coupled to
 another single-band superconductor by the Josephson effect. The two bands are assumed to have separated
 Fermi surfaces in $\vec{k}$-space and are coupled only through electron-electron interaction. We shall assume for
 simplicity that the superconducting order parameters on all Fermi surfaces have $s$-symmetry, although most of
 our results can be generalized to order-parameters with other symmetries as well. We start with the
 phenomenological Ginzburg-Landau (GL) free energy density of the two band superconductor.
 \begin{equation}
 \label{free0}
 F_0(\psi_1,\psi_2)=\alpha_1(T)|\psi_1|^2+K_1|\vec{D}\psi_1|^2+\beta_1|\psi_1|^4+\alpha_2(T)|\psi_2|^2
 +K_2|\vec{D}\psi_2|^2+\beta_2|\psi_2|^4-J[\psi_1^*\psi_2+\psi_2^*\psi_1]
 \end{equation}
 where $\psi_{1(2)}$ is the superconducting order parameter for band 1(2) and
 $\vec{D}=-i[\nabla-2ei\vec{A}/\hbar c]$. In general the two superconducting bands are coupled by an
 {\em internal} Josephson-coupling term $\sim J$ as a result of electron-electron interaction. Writing
 $\psi_{1(2)}=|\psi_{1(2)}|e^{i\theta_{1(2)}}$ and minimizing the energy with respect to $\theta_{1(2)}$ in the
 absence of magnetic field it is easy to see that $\psi_1$ and $\psi_2$ are of the same sign
 $(\theta_1=\theta_2 (mod 2\pi))$ if $J>0$, and are of opposite sign ($\theta_1=\theta_2+\pi (mod 2\pi))$ if
 $J<0$. The question is whether there exists any non-trivial physical consequences associated with this relative
 sign, in particular when $J<0$ and $\psi_1\sim-\psi_2$?

    In the following we shall show that the spontaneous time-reversal symmetry breaking occurs at the Josephson
 junction between the two-band superconductor and another single-band $s$-wave superconductor when the sign of $J$
 is negative. To be concrete we assume that the single-band superconductors is located at the left side ($x<0)$ of
 the Josephson junction, and the two-band superconductor is located on the right ($x>0)$. The two superconductors
 are weakly coupled by Josephson effect, and the total free energy density of the system is
 $F=F_0\theta(x)+F_s\theta(-x)+F_J$, where
  \begin{equation}
  \label{frees}
  F_s(\psi_s)=\alpha_s(T)|\psi_s|^2+K_s|\vec{D}\psi_s|^2+\beta_s|\psi_s|^4
  \end{equation}
  is the usual Ginsburg-Landau free energy for the single-band $s$-wave superconductor, and
  \begin{equation}
  \label{freej}
  F_J=-\left(T_1[\psi_1^*\psi_s+\psi_s^*\psi_1]+T_2[\psi_2^*\psi_s+\psi_s^*\psi_2]\right)\delta(x)
  \end{equation}
 is the Josephson coupling between the two superconductors. $T_{1(2)}$ represents the coupling of the single-band
 superconductor to the two separate bands. We note that $T_1$ and $T_2$ are both positive according to the
 perturbation theory in the tunnelling matrix elements between the two superconductors. The relative sign between
 $\psi_1$ and $\psi_2$ is ``unknown" to the single-band superconductor in the Josephson effect.

  The non-trivial effect associated with the Josephson junction can be seen by minimizing the free energy of the
  system with respect to the phases of the superconductors, assuming that the amplitudes of the order parameters
  are constants. In the absence of magnetic field the GL free energy density with phase variables only is
  \begin{eqnarray}
  \label{freep}
   F & \sim & \bar{F}+\theta(x)\left(-2\tilde{J}\cos(\theta_1-\theta_2)+\tilde{K}_1(\nabla\theta_1)^2
  +\tilde{K}_2(\nabla\theta_2)^2\right)  \\ \nonumber
  & & -2\delta(x)\left(\tilde{T}_1\cos(\theta_1-\theta_s)+\tilde{T}_2\cos(\theta_2-\theta_s)\right)
  +\tilde{K}_s(\nabla\theta_s)^2
  \end{eqnarray}
  where $\bar{F}$ is the part of free energy density which is independent of $\theta$'s. $\tilde{\theta}_{1(2)}$
  is defined at $x\geq0$ and $\theta_s$ is defined as $x\leq0$. $\tilde{J}=J|\psi_1||\psi_2|$,
  $\tilde{T}_{1(2)}=T_{1(2)}|\psi_s||\psi_{1(2)}|$ and $\tilde{K}_{\nu}=K_{\nu}|\psi_{\nu}|^2$, $\nu=1,2,s$.
  We shall take
  $\theta_s(x=0)=0$ in the following. This is allowed because the overall phase of the system is a pure gauge.
  First we note that for $J>0$ the phase configuration which minimizes the above free energy is simply
  $\theta_1=\theta_2=\theta_s=0$ and there is no non-trivial effect associated with the Josephson junction. The
  situation becomes different for $J<0$ where the phases are ``frustrated" because of sign difference between $J$
  and $T_{1(2)}$. To see what could happen we note that the system is similar to a system of two classical spins
  $A$ and $B$ that are antiferromagnetically coupled and are put under a weak magnetic field in $\hat{x}$
  direction representing the Josephson coupling of the system to the single-band superconductor. If the coupling
  of spin $A$ to magnetic field is much stronger than that of spin $B$ ($T_1>>T_2$), spin $A$ will be allied to the
  magnetic field with spin $B$ remaining anti-parallel to spin $A$ in the ground state, i.e. $\theta_1=0,\theta_2=
  \pi$ in the corresponding Josephson junction problem. The converse ($\theta_2=0,\theta_1=\pi$) is true if
  $T_2>>T_1$. This is the first possible state. Notice however that if the couplings of the two spins to the
  magnetic field are similar, there will be no preferred spin to the magnetic field and a second type of state
  where the two spins take angles $\theta_{1(2)}\sim\pm\pi/2\mp\delta\theta$ will be formed (see Fig.1).

  \begin{figure}
  \includegraphics[width=7.0cm, angle=0]{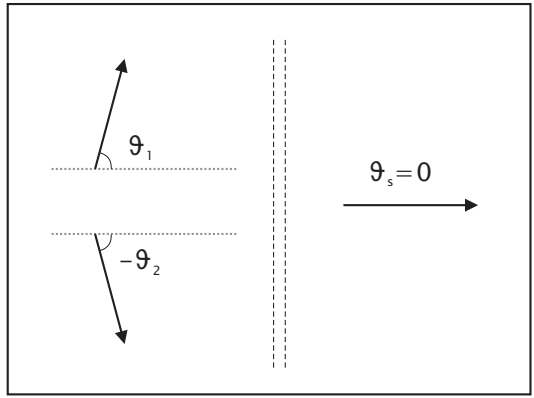}
  \caption{ A schematic representation of canted state in the effective classical
  spin model for the Josephson junction between a two band superconductor and another single-band superconductor.
  The coupling between the two spins ($\theta_1,\theta_2$) are antiferromagnetic ($J<0$) but they are both
  coupling ferromagnetically to a magnetic field along $\hat{x}$-direction ($\theta_s=0$)}
  \end{figure}

  The free energy\ (\ref{freep}) cannot be minimized exactly to obtain the two types of phase structures. We shall
  treat
  the free energy approximately in the following by writing $\theta_{1}=\theta_0+\tilde{\theta}_1$,
  $\theta_2=\theta_0-\pi+\tilde{\theta}_2$ and expand the free energy to order $\tilde{\theta}_{1(2)}^2$.  The
  approximation can be justified in the limit of weak-Josephson coupling as we shall see later.

   Minimizing the resulting approximate free energy $F(\theta_0,\tilde{\theta})$ with respect to
 $\tilde{\theta}$'s we obtain
  \begin{eqnarray}
  \label{alpha}
  \tilde{J}(\tilde{\theta}_1-\tilde{\theta}_2)+\tilde{K}_1\nabla^2\tilde{\theta}_1 & = &
  \tilde{T}_1\delta(x)(\sin\theta_0+\cos\theta_0\tilde{\theta}_1), \\ \nonumber
  \tilde{J}(\tilde{\theta}_2-\tilde{\theta}_1)+\tilde{K}_2\nabla^2\tilde{\theta}_2 & = &
  -\tilde{T}_2\delta(x)(\sin\theta_0+\cos\theta_0\tilde{\theta}_2), \\ \nonumber
  \tilde{K}_s\nabla^2\theta_s & = &  -\delta(x)\left(\tilde{T}_1(\sin\theta_0+\cos\theta_0\tilde{\theta}_1)-
   \tilde{T}_2(\sin\theta_0+\cos\theta_0\tilde{\theta}_2)\right).
  \end{eqnarray}

    Solving these equations at $x\neq0$ we obtain $\tilde{\theta}_{1(2)}=\alpha_{1(2)}e^{-x/\lambda}+\beta_0x$ and
    $\theta_s=\beta_sx$ where $\tilde{K}_1\alpha_1=-\tilde{K}_2\alpha_2$ and
    ${1\over\lambda^2}=|\tilde{J}|{(\tilde{K}_1+\tilde{K}_2)\over\tilde{K}_1\tilde{K}_2}$.

    Matching the boundary condition at $x=0$ we also obtain
  \begin{eqnarray}
  \label{bc}
  J_1=\tilde{K}_1(\beta_0-{\alpha_1\over\lambda}) & = &
  \tilde{T}_1(\sin\theta_0+\cos\theta_0\alpha_1),  \\  \nonumber
  J_2=\tilde{K}_2(\beta_0-{\alpha_2\over\lambda}) & = &
  -\tilde{T}_2(\sin\theta_0+\cos\theta_0\alpha_2),  \\  \nonumber
  \tilde{K}_s\beta_s & = & J_1+J_2.
  \end{eqnarray}
   The first two equations give the tunnelling currents flowing from band one (two) to the single-band
  superconductor, respectively. The third equation expresses total current conversation across the Josephson
  junction.

  We shall concentrate on the ground state solution where the Josephson junction does not introduce any bulk energy
 cost. In this case there is no net current flowing through the system and $\beta_0=\beta_s=0$. Solving Eq.\
 (\ref{bc}) we find that only one solution $\cos\theta_0=sgn(\tilde{T}_1-\tilde{T}_2)$ exists at
 $|\tilde{T}_1-\tilde{T}_2|>\sigma_1$, where $\sigma_1={\lambda\tilde{T}_1\tilde{T}_2(\tilde{K}_1+\tilde{K}_2)
 \over\tilde{K}_2\tilde{K}_1}$ whereas two possible solutions $\cos\theta_0=\left(sgn(\tilde{T}_1-\tilde{T}_2),
 {(\tilde{T}_1-\tilde{T}_2)\over\sigma_1}\right)$ exist at $|\tilde{T}_1-\tilde{T}_2|<\sigma_1$. The true solution
 at $|\tilde{T}_1-\tilde{T}_2|<\sigma_1$ is the one with lower energy. Comparing the energies $F(\theta_0,
 \tilde{\theta})$ of the two states we find
  \begin{eqnarray}
  \label{scos}
  \cos\theta_0 & = & sgn(\tilde{T}_1-\tilde{T}_2) \,\,\,\,\,\, (|\tilde{T}_1-\tilde{T}_2|>\sigma)
  \\ \nonumber
   & = & {(\tilde{T}_1-\tilde{T}_2)\over\sigma_1} \,\,\,\,\,\,\,\,\,\,\,\, (|\tilde{T}_1-\tilde{T}_2|<\sigma)
  \end{eqnarray}
  where
  $\sigma=min(\sigma_1,\sigma_2)$, $\sigma_2={\lambda(\tilde{K}_2\tilde{T}_1+\tilde{K}_1\tilde{T}_2)^2
  \over(\tilde{K}_1+\tilde{K}_2)\tilde{K}_2\tilde{K}_1}$.

 In the first case $\theta_0=0$ or $\pi$, which corresponds to the first type of solution in the classical spin
 problem. The solution respects time-reversal symmetry and we shall call it $TRI$ state in the following.

    There are two degenerate solutions in the second case corresponding to $\theta_0\gtrless0$ which are
 time-reversal pairs ($\psi_{1(2)}\rightleftharpoons\psi^*_{1(2)}$). The solution breaks time reversal symmetry
 and we shall call it the $TRB$ state. The corresponding $\alpha_{1(2)}$ is equal to zero in the $TRI$ state, and
 is of order $\sin\theta_0\tilde{T}_{1(2)}\lambda/\tilde{K}_{1(2)}$ in the $TRB$ state, which is much less than
 $\theta_0$ in the limit of weak-Josephson coupling $\tilde{T}_{1(2)}\lambda/\tilde{K}_{1(2)}<<1$, justifying our
 approximate treatment of free energy.

  It is interesting to note that although the net Josephson current passing through the tunnelling barrier
 is zero in the ground state, the currents $J_1$ and $J_2=-J_1$, which represent current passing from band $1(2)$
 of the two-band superconductor to the single-band superconductor, are nonzero in the $TRB$ state. Correspondingly
 there is also a nonzero current $J_{12}\sim\tilde{J}\sin(\theta_1-\theta_2)$ passing from band one to band two
 in the $TRB$ state.

     Thus the $TRB$ state is characterized by a novel current ``loop" through the Josephson function. A
    current flows from band one/two of the two-band superconductor to the single-band superconductor through the
    Josephson junction, and flows back to band two/one of the two-band superconductor. The current flow from band
    two/one to band one/two inside the two-band superconductor to complete the current loop. The current loop
    we see here is not a current loop circulating in real space, but a current loop in $\vec{k}$-space,
    if we envision the two bands as occupying different parts of the $\vec{k}$-space.

      It is also straightforward to see from Eq.\ (\ref{bc}) that the parameter space where the $TRB$ state exists
     is enlarged when there is a net current flowing across the Josephson junction ($\beta_0,\beta_s\neq0$). This
     is not surprising since a finite current through the system breaks time-reversal symmetry. The current also
     removes the degeneracy of the two $TRB$ solutions with $\theta_0\lessgtr0$.

    The presence of non-trivial phase structure leads to non-trivial electron surface states on the surface of
   the Josephson junction. To study these surface states we consider the ground state and analyze the corresponding
   Bogoliubov-de Gennes (BdG) equation,
    \begin{eqnarray}
    \label{bdg}
    \epsilon_n^{(i)}u_n^{(i)}(\vec{x}) & = & \hat{H}_o^{(i)}u_n^{(i)}(\vec{x})+\Delta^{(i)}(\vec{x})
    v_n^{(i)}(\vec{x}) \\  \nonumber
    \epsilon_n^{(i)}v_n^{(i)}(\vec{x}) & = & -\hat{H}_o^{(i)}v_n^{(i)}(\vec{x})+\Delta^{(i)*}(\vec{x})
    u_n^{(i)}(\vec{x})
    \end{eqnarray}
    where $i=1,2$ are the band indices, $\hat{H}_o^{(i)}$ gives the single-particle band structure for band $i$ and
    $\Delta^{(i)}(\vec{x})$ is the corresponding superconducting order parameter. We have assumed that the two
    bands are independent of each other and are coupled implicitly only through the superconductor
    order parameter in writing down Eq.(\ref{bdg}). In particular, $\Delta^{(1)}\sim+(-)\Delta^{(2)}$ in
    the bulk superconductor if $J>(<)0$. The Josephson function can be modeled by superconducting order parameters
    of the form $\Delta^{(i)}(\vec{x})\sim\Delta_0^{(i)}e^{i\theta_{i}(x)}$ at $x>0$ and $\Delta(x)=\Delta_s$ at
    $x<0$, where $\Delta_0^{(i)}$ and $\Delta_s$ are real and positive. We shall show in the following that
    non-trivial electronic surface state exists in the Josephson junction, with structures depending strongly
    on the phases of the superconducting order parameters.

      The BdG equations for electronic states close to the Fermi surface can be solved in the WKBJ
    approximation\cite{hu} where we write $u(v)_n^{(i)}(\vec{x})\sim e^{i\vec{k}_F.\vec{x}}\tilde{u}
    (\tilde{v})_n^{(i)}(\vec{x})$, where $\tilde{u}(\tilde{v})_n^{(i)}(\vec{x})$ are slowly varying functions of
    $\vec{x}$ (on the scale of $k_F^{-1}$) satisfying the Andreev equations,
    \begin{eqnarray}
    \label{and}
    \epsilon_n^{(i)}\tilde{u}_n^{(i)}(\vec{x}) & = & -i(\vec{v}_F^{(i)}.\nabla)\tilde{u}_n^{(i)}(\vec{x})
    +\Delta^{(i)}(\vec{x})\tilde{v}_n^{(i)}(\vec{x}) \\  \nonumber
    \epsilon_n^{(i)}\tilde{v}_n^{(i)}(\vec{x}) & = & i(\vec{v}_F^{(i)}.\nabla)\tilde{v}_n^{(i)}(\vec{x})+
    \Delta^{(i)*}(\vec{x})\tilde{u}_n^{(i)}(\vec{x}),
    \end{eqnarray}
    where $\vec{v}_F^{(i)}$ is the Fermi velocity of the band $i$ electrons
with momentum $\vec{k}_F^{(i)}$. We shall
    consider the weak-Josephson coupling limit ($\tilde{\theta}_{1(2)}<<\theta_0$) in the following, so that
    $\theta_1(x)\sim\theta_0$ and $\theta_2(x)\sim\theta_0-\pi$. The bound states are given by solutions of form
    $\tilde{u}(\tilde{v})_n(\vec{x})=\tilde{u}(\tilde{v})_0e^{-\gamma_+x}$ for $x>0$ and
    $\tilde{u}(\tilde{v})_n(\vec{x})=\tilde{u}(\tilde{v})_0e^{\gamma_-x}$ for $x<0$.
    Substituting these into Eq.\ (\ref{and}), we obtain the self-consistent equations
    \begin{eqnarray}
    \label{sol1}
    \epsilon_0^{(i)2}=\Delta^{(i)2}_0(1-x^{(i)2}) & = & \Delta^2_s(1-y^{(i)2}) \\  \nonumber
    {\epsilon_0^{(i)}/\Delta^{(i)}_0-ix^{(i)}\over\epsilon_0^{(i)}/\Delta_s+iy^{(i)}} & = & e^{i\theta_{i}}
    \end{eqnarray}
    where $x^{(i)}=\vec{v}_F^{(i)}.\hat{x}(\gamma^{(i)}_+/\Delta^{(i)}_0)$ and $y^{(i)}=\vec{v}_F^{(i)}.\hat{x}
    (\gamma^{(i)}_-/\Delta_s)$. Notice that $x^{(i)}$ and $ y^{(i)}$ must have the same sign in this
    representation and changing sign of $x(y)^{(i)}$ corresponds to changing $\vec{v}_F\rightarrow-\vec{v}_F$.

    It is easy to see that if $x^{(i)}, y^{(i)}$ is a solution of Eq.\ (\ref{sol1}) with energy $\epsilon_0^{(i)}$,
    then $-x^{(i)}, -y^{(i)}$ is a solution with energy $-\epsilon_0^{(i)}$. $-x^{(i)}, -y^{(i)}$ is also a
    solution with energy $\epsilon_0^{(i)}$ with $\theta_i\rightarrow-\theta_i$. Therefore it is sufficient to consider the range
    $\pi>\theta_i>0$. Solving Eq\ (\ref{sol1}) we find that solutions where $x^{(i)}$ and $ y^{(i)}$ have the same
    sign exists only when $\cos\theta_i<min (\Delta_0^{(i)}/\Delta_s,\Delta_s/\Delta_0^{(i)})$, with
    \begin{eqnarray}
    \label{sol2}
    x^{(i)} & = & {\Delta_s\cos\theta_i-\Delta_0^{(i)}\over D^{(i)}}sgn(\epsilon_0^{(i)}),  \\  \nonumber
    y^{(i)} & = & {\Delta_0^{(i)}\cos\theta_i-\Delta_s\over D^{(i)}}sgn(\epsilon_0^{(i)}),  \\  \nonumber
    \epsilon_0^{(i)} & = & \pm{\Delta^{(i)}_0\Delta_s\sin\theta_i\over D^{(i)}}.
    \end{eqnarray}
    where $D^{(i)}=\sqrt{\Delta_s^2+\Delta_0^{(i)2}-2\cos\theta_i\Delta_s\Delta_0^{(i)}}$. Notice that there exists
    one solution for each value of Fermi momentum $\vec{k}_F^{(i)}=(k_{Fx}^{(i)}, k_{Fy}^{(i)},k_{Fz}^{(i)})$.
    Thus a finite density of states exist at the Josephson junction in general.

    We now analyze the solutions as a function of $\theta_i$. First we note that bound state solutions do not exist
    when $\theta_i=0$, i.e. when $J>0$ and the system is not frustrated. For $J<0$ and $\theta_i=\pi$, we find that bound state solutions with $\epsilon^{(i)}_0=0$
    exist at both $x(y)^{(i)}\lessgtr0$ channels. In the corresponding Josephson junction problem, bound
    state solutions with zero energy exist in the $TRI$ state in the band which is out-of-phase with the
    single-band superconductor, but there is no bound state solution for the band which is in-phase with the
    single-band superconductor. Since time-reversal symmetry is preserved in the $TRI$ state, the
    $x(y)^{(i)}\lessgtr0$ states have the same bound state energy $\epsilon^{(i)}_0=0$.

  The bound state structure is much richer in the $TRB$ state which breaks time-reversal symmetry. In this case,
 the structure of the bound states depend on the value of $\theta_0$ (Recall $\theta_1\sim\theta_0$ and
 $\theta_2\sim\theta_0-\pi$). First we consider $\theta_0<\pi/2$. For small $\theta_0$ such that
 $\cos\theta_0>min(\Delta_0^{(1)}/\Delta_s,\Delta_s/\Delta_0^{(1)})$, bound state solutions exist only in
 band two. The bound states exist in pairs with energies $\pm\epsilon_0^{(2)}\neq0$, corresponding to
 time-reversal pairs $x(y)^{(2)}\lessgtr0$. For larger $\theta_0$ such that $\cos\theta_0<min(\Delta_0^{(1)}
 /\Delta_s, \Delta_s/\Delta_0^{(1)})$, bound state solutions exist in both bands in
  time-reversal pairs with energies $\pm\epsilon_0^{(1,2)}\neq0$. Similar results occur for $\theta_0>\pi/2$
  where bound states exist only in band one if $-\cos\theta_0>min(\Delta_0^{(2)}/\Delta_s,\Delta_s/\Delta_0^{(2)})$
  and exist in both bands if $-\cos\theta_0<min(\Delta_0^{(2)}/\Delta_s,\Delta_s/\Delta_0^{(2)})$.

   Now the experimental observation of this bound state is discussed. First, the $dI/dV$ curve for the Josephson
  tunneling should show the peak at the bound state energy within the gap. A zero-bias peak would exists in the
  $TRI$ state, and is split into two peaks centered at finite energies in the $TRB$ state. The local probe
  such as the STS can also be used to detect these bound states. The detection of the persistent Josephson current
  loop in $\vec{k}$-space in the $TRB$ state will be a challenge. The standard prove for the time-reversal
  symmetry breaking is the Kerr rotation. With the spin-orbit interaction, the finite spin density is expected at
  the Josephson junction, but the details of the analysis depends on band structure which we have not
  undertaken in this paper.

  We emphasize that the effects discussed in this paper are general effects associated with frustrated phase
 structures in superconductor Josephson junctions, and is not restricted to $s$-wave superconductor, or Josephson
 junction involving single-band and two-band superconductors. This idea of the frustration can be generalized to
 single multi-band superconductors with three or more bands coupled {\em via internal} Josephson effect. In this
 case, we can construct the effective XY-spin model with positive or negative exchange interactions between pairs of the
 order parameters, and when there exists relative angle(s) different from $0$ or $\pi$, time-reversal symmetry is
 broken spontaneously. This mechanism is likely to be active in the superconductors with rather complex band
 structure such as the heavy fermion systems, where many sheets of the Fermi surface contribute to the pairing.

To summarize, we have studied the Josephson junction between the
two-gap superconductor and the conventional s-wave superconductor.
When the relative sign of the two gaps are negative, the Josephson
coupling introduces the frustration, which can lead to the
time-reversal symmetry breaking near the junction. This results in
the bound state within the gaps, which an be detected by the
$dI/dV$ characteristics or STS, which can offers an experimental
test of the relative phase of the two gaps.

 The authors are grateful to Patrick A Lee for fruitful discussions. This work was
supported in part by
 Grant-in-Aids (Grant No. 15104006, No. 16076205, and No. 17105002) and NAREGI Nanoscience Project from the
 Ministry of Education, Culture, Sports, Science, and Technology. TKN also
 acknowledge support by HKUGC through grant CA05/06.Sc04.

\end{document}